\documentclass{iau}
\usepackage{graphicx}

\title[Molecular Gas and Star Formation in Voids] 
{Molecular Gas and Star Formation in Void Galaxies}

\author[M.~Das et al.]   
{M.~Das$^1$, T.~Saito$^2$, D.~Iono$^2$, M.~Honey$^1$ 
 \and S.~Ramya$^3$}

\affiliation{$^1$Indian Institute of Astrophysics, Banaglore, India \\
email: {\tt mousumi@iiap.res.in} \\[\affilskip]
$^2$Department of Astronomy, The University of Tokyo, Tokyo 113-0033 \\[\affilskip]
$^3$Shanghai Astronomical Observatory, Shanghai }
\pubyear{2014}
\volume{308}  
\pagerange{}
\jname{The Zeldovich Universe: Genesis and Growth of the Cosmic Web}
\editors{Rien van de Weygaert, Sergei Shandarin, Enn Saar \& Jaan Einasto}
\begin{document}
\maketitle

\begin{abstract}

We present the detection of molecular gas using CO(1--0) line emission and followup H$\alpha$ imaging 
observations of galaxies located in nearby voids. The CO(1--0) observations were done using the 45m telescope 
of the Nobeyama Radio Observatory (NRO) and the
optical observations were done using the Himalayan Chandra Telescope (HCT). Although void galaxies lie in the most 
underdense parts of our universe, a significant fraction of them are gas rich, spiral galaxies that show signatures 
of ongoing star formation. Not much is known about their cold gas content or star formation properties. In
this study we searched for molecular gas in five void galaxies using the NRO. The galaxies
were selected based on their relatively higher IRAS fluxes or H$\alpha$ line luminosities. CO(1--0) emission was
detected in four galaxies and the derived molecular gas masses lie between $(1 - 8)\times10^{9}~M_{\odot}$. The
H$\alpha$ imaging observations of three galaxies detected in CO emission indicates ongoing star formation and the derived
star formation rates vary between from 0.2 - 1.0~M$_{\odot}~yr^{-1}$, which is similar to that observed
in local galaxies. Our study shows that although void galaxies reside in underdense 
regions, their disks may contain molecular gas and have star formation rates similar to galaxies
in denser environments.

\end{abstract}

\firstsection 
\section{Introduction}

Voids contain a sparse but significant population of galaxies that are usually small, gas rich, late type galaxies 
(kreckel etal. 2012). The smaller voids are dominated by low surface brightness (LSB) dwarfs and irregular 
galaxies (karachentsev etal. 1999) but the larger voids also have a population of relatively bright galaxies
that are often blue in color. These galaxies have ongoing star formation and are often interacting with companion
galaxies in pairs or small groups along filaments in the voids (Beygu et al. 2013). Many questions remain 
regarding star formation in void environments; what is its nature - is it sporadic or continuous, what drives 
it and how is the star formation related to the location of the galaxies with respect to the filaments, walls and void 
interiors~?   

One of the key elements for supporting star 
formation in galaxies is the presence of molecular hydrogen ($H_2$) gas. Although neutral hydrogen has been 
both detected and mapped in several voids, not much is known about the distribution of $H_2$ gas in void galaxies. 
There have been two studies that have detected CO emission and estimated molecular gas masses in a total of five 
void galaxies (Sage et al. 1997; Beygu et al. 2013). These results indicate that the $H_2$ gas masses in void galaxies are 
comparable to those found in nearby star forming systems. In this study we searched for molecular gas in 
void galaxies to obtain a larger sample of such $H_2$ rich galaxies and carried out followup H$\alpha$ imaging observations 
of some of the detected galaxies. Our main motivation was to understand how the cold gas masses relate to the star 
formation properties of these systems. In the following sections we present our observations, results and discuss their 
implications. For all distances we have used $H_{0}~=~73~km~s^{-1}~Mpc^{-1}$ and $\Omega~=~0.27$.

\begin{figure}[b]
\begin{center}
 \includegraphics[width=2.5in]{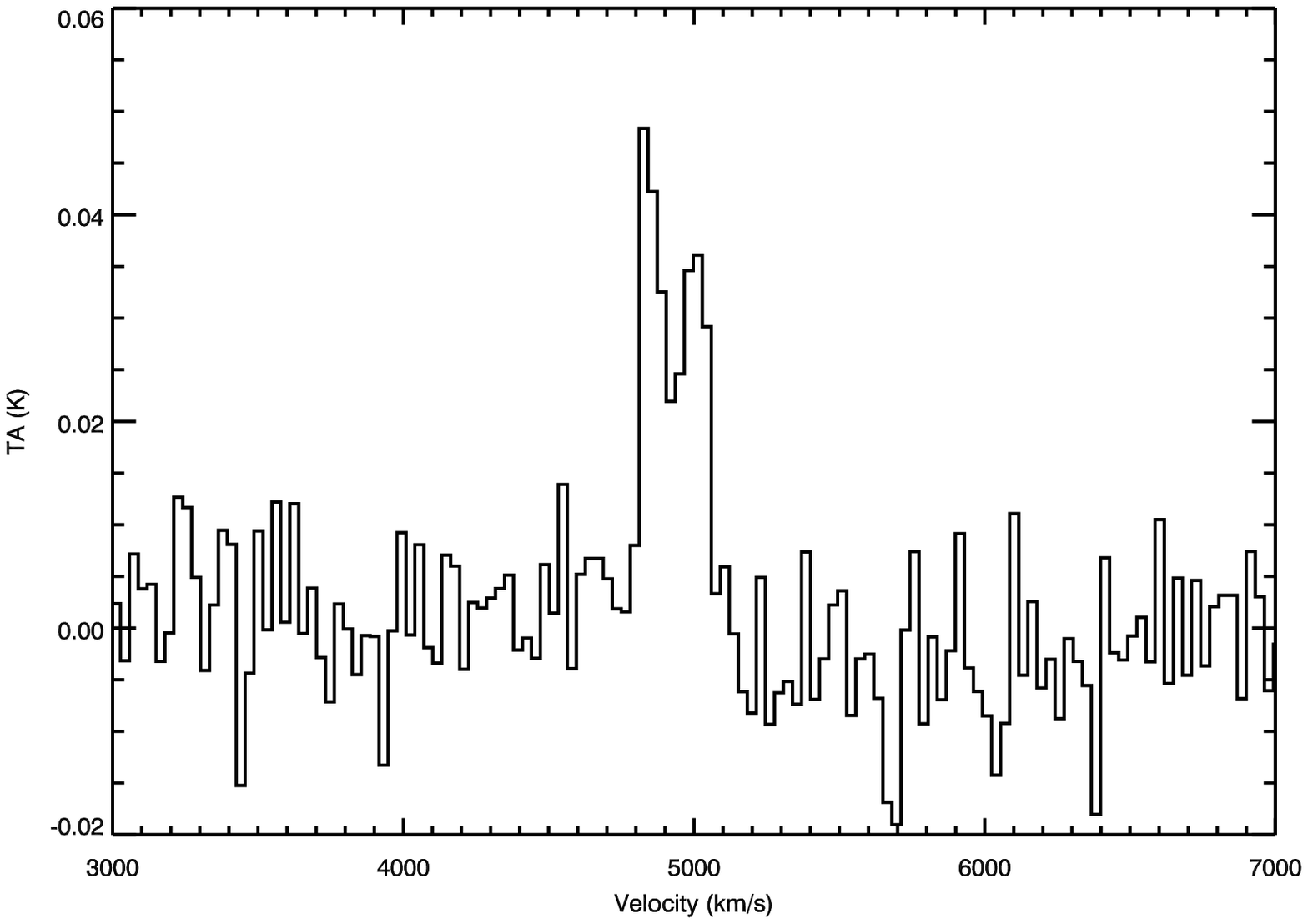} 
 \includegraphics[width=2.5in]{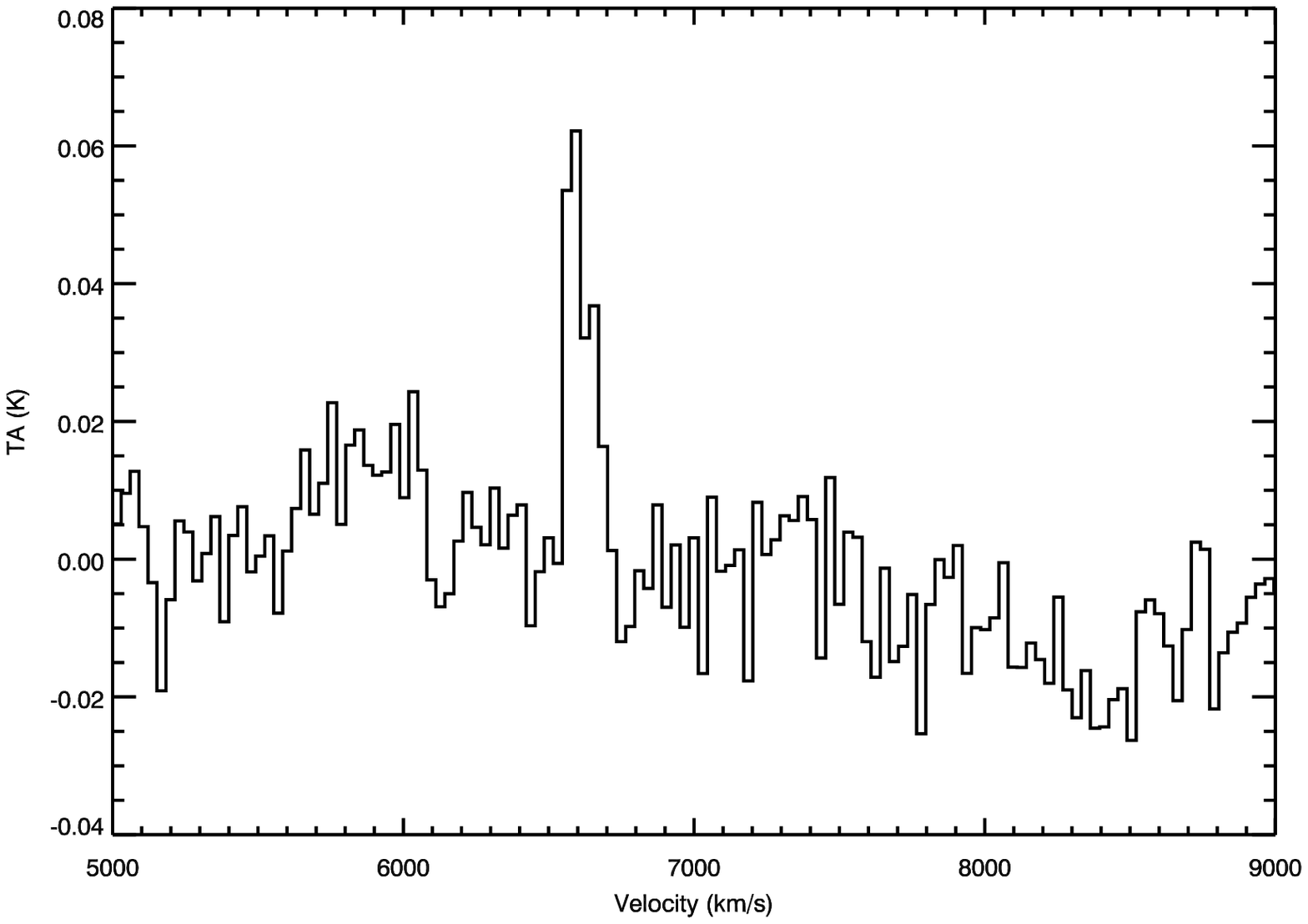}
 \caption{(a)~Figure on the left shows the CO(1--0) line emission detected from the galaxy SBS~1325+597 that lies in the void
Ursa Minor~I. The line has a distinctive double horned profile indicative of a rotating disk and the peak separation is  
approximately 200~km~s$^{-1}$. (b)~Figure on the right shows the CO(1--0) line emission detected from the void galaxy 
SDSS~153821.22+331105.1. The gas is concentrated in the center of the galaxy and has a line width of $\sim$100~km~s$^{-1}$.}
   \label{fig1}
\end{center}
\end{figure}

\vspace{-1mm}
\section{Sample galaxies and observations}

Our initial sample comprised of 12 galaxies that were selected based of their relatively
high infra-red fluxes or high star formation rates (kreckel et a. 2012; Cruzen et al. 2002; Szomoru et al. 1996).
However due to weather conditions we were able to finally observe only five galaxies
from this sample and they are listed in Table~1. Three galaxies have been observed in HI by Kreckel et al. (2012) 
in the Void Galaxy Survey (VGS) (SBS~1325+597, SDSS 143052.33+551440.0, SDSS 153821.22+331105.1) and the remaining 
two galaxies have been observed in an earlier HI survey of the Bootes void by Szomoru et al. (1996). All the galaxies 
have SDSS data.

The $^{12}CO (J=1-0)$ 
emission observations were carried out using the 45~m Nobeyama Radio Telescope during 14 - 25 April, 2013. At the CO rest 
frequency of 115.271204 GHz, the half-power beam width (HPBW) was 15$^{\prime\prime}$ and the main beam efficieny was 
about 30\%. The on source time for the first four galaxies varied between 1 to 1.5 hours; due to poor weather conditions 
SBS~1428+529 was observed for only 25 minutes. We used the one beam (TZ1), dual polarization, double sideband receiver (TZ) and 
the digital FX-type spectrometer SAM45, that has a bandwidth of 4~GHz (Nakajima et al. 2008). Typical system temperatures were 
160 - 260~K. The pointing accuracy was about 2$^{\prime\prime}$-4$^{\prime\prime}$. Only data with a wind velocity less than 
5 km~s$^{-1}$ were used and data with winding baselines were flagged. The data was analysed using NRO calibration tool NEWSTAR.

The H$\alpha$ observations were done using the Himalayan Faint Object Spectrograph 
Camera (HFOSC) which is mounted on the 2m Himalayan Chandra Telescope (HCT) and were carried out on 2014 April 11 \& 25.
For SBS~1325+597 the redshift is 0.0165, so the H$\alpha$ filter (band width $\sim$~500~$\AA$) was used to get the 
H$\alpha$ line emission. For SDSS 143052.33+551440.0 we used the narrow H$\alpha$ filter (band width $\sim$~100~$\AA$). The galaxy 
SDSS J153821.22+331105.1 is at a redshift of $z=0.022$ and the H$\alpha$ line is shifted to 6714~$\AA$. Hence we used the narrow band 
[SII] filter (band width $\sim$~100~$\AA$ and centered around 6724~$\AA$) for this galaxy. To obtain the continuum subtracted H$\alpha$ 
images we also obtained broad band images with the $R$ filter centered around the H$\alpha$ 
line. The bias frames and twilight flats were used for preprocessing of the images. The data reduction was done 
using the standard packages available in IRAF\footnote{Image Reduction \& Analysis Facility Software distributed 
by National Optical Astronomy Observatories, which are operated by the Association of Universities for Research in 
Astronomy, Inc., under co-operative agreement with the National Science Foundation}. The images were corrected for cosmic rays, 
aligned and corrected for point spread function variations. Flux calibration was done using the 
spectrophotometric standard star HZ44. The H$\alpha$ fluxes are listed in Table~1.

\begin{table}
  \begin{center}
  \caption{Observed galaxies, molecular gas masses and star formation rates}
  \label{tab1}
 {\scriptsize
  \begin{tabular}{|l|c|c|c|c|c|c|}\hline 
{\bf Galaxy} & {\bf D$_{L}$} & {\bf Redshift} & {\bf CO flux}         & {\bf Molecular Gas } & {H$\alpha$ Flux ($10^{-13}$} & {\bf SFR }\\
             & {\bf Mpc}     &   & {\bf (K~km~s$^{-1}$)} & {\bf Mass($10^{9}~M_{\odot}$)} & {~ergs~s$^{-1}$cm$^{-2}$)} & {\bf ($M_{\odot}yr^{-1}$)}\\
\hline
 SBS 1325+597 &  70.4        & 0.0165         &  10.7$\pm$0.2         & 1.5$\pm$0.03    &  0.4  &  0.20 \\
 SDSS 143052.33+551440.0 & 76.6 & 0.0176      &   7.0$\pm$0.2         & 1.1$\pm$0.03    &  1.2  &  0.60 \\
 SDSS 153821.22+331105.1 & 97.6 & 0.0220      &   6.4$\pm$0.2         & 1.7$\pm$0.05    &  1.2  &  1.02 \\
 CG 598                  & 248.0 & 0.0575     &   5.2$\pm$0.1         & 8.5$\pm$0.10    &  ....  &  ..... \\
 SBS 1428+529            & 191.0 & 0.0445     &   $<~$0.6             &  $<~$0.6        &  ....  &  ..... \\ 
\hline
  \end{tabular}
  }
 \end{center}
 \scriptsize{
 {\it Notes:}\\
  $^1$ For SBS 1428+529 there was no CO(1--0) detection and no H$\alpha$ image could be obtained. Upper limits for the molecular 
gas mass were obtained from the noise which was 0.0024~k and assuming a typical linewidth of 250~km/s.\\
  $^2$ For CG 598 no H$\alpha$ image could be obtained.}
\end{table}

\vspace{-1mm}
\section{Results}

\noindent
{\bf 1.~Molecular gas detection~:~}We have detected $^{12}CO (J=1-0)$ emission from four of the five sample galaxies that we 
observed (Table~1). The non-detection in SBS~1428+529 could partly be due to the short duration of the scan, which was limited by bad 
weather. Of the four detections, SBS~1325+597 has the most striking line profile; it has a double horned structure indicating a rotating 
disk of molecular gas (Figure~1a). The velocity separation of the peaks is $\sim200~kms^{-1}$; assuming a disk inclination of 59.3$^{\circ}$ the 
disk rotation is 116~km~s$^{-1}$. This is similar to the HI rotation speed from Kreckel et al. (2012). In SDSS 153821.22+331105.1 the 
gas is centrally peaked (Figure~1b); probably driven into the center by the bar in the galaxy. In the other two galaxies 
(SDSS 143052.33+551440.0 and CG~598) the CO line profile is slightly off 
center from the systemic velocities of the galaxies, which suggests that their gas disks are disturbed, possibly due to interaction with 
a companion galaxy (Das et al. 2014, in preparation).    \\
{\bf 2.~Molecular masses~:~} The CO fluxes in K~km~s$^{-1}$ were converted to Jy~km/s using a conversion factor (Jy/K) of 2.4. The CO line 
luminosity was determined using the relation $L_{CO}~=~3.25\times10^{7}(S_{CO}\Delta~V/Jykms^{-1})(D_{L}/Mpc)^{2}(\nu_{res})^{-2}(1+z)^{-1}$  
and the molecular gas mass was estimated using the relation $M(H_{2})~=~[4.8~L_{CO}(K~kms^{-1}$)] (Solomon \& van den Bout 2005). The molecular 
gas masses lie in the range $(1-8)\times10^{9}~M_{\odot}$ which is comparable to that observed from bright galaxies in denser environments.\\
{\bf 3.~Comparison with previous detections and HI masses~:~}Our molecular gas are similar to that obtained for earlier studies of void 
galaxies by (Sage et al. 1997,Beygu et al. 2013) that lie in the range $10^{8} - 10^{9}$~M$_{\odot}$. Although our study and previous 
detections indicate relatively large $H_2$ gas masses and a high detection rate, it must be remembered that the sample was biased towards 
star forming galaxies and those with high FIR fluxes. The molecular gas masses are comparable to the HI masses of these galaxies 
(Kreckel et al. 2012; Szomoru et al. 1996).\\
{\bf 4.~H$\alpha$ fluxes and star formation rates (SFR)~:~}The SFRs were calculated from the H$\alpha$ fluxes using the kennicutt formula 
SFR~=~L(H$\alpha$)/1.26$\times$10$^{41}$~ergs~s$^{-1}$ (Table~1). In SBS~1325+597 the H$\alpha$ is distributed on either side of the galaxy
nucleus, possibly in a ring. The distrbution matches the CO line profile (Figure~1a) which indicates a ring like configuration for the 
molecular gas.  In SDSS~143052.33+551440.0 the H$\alpha$ emission is concentrated about the nucleus. In SDSS 153821.22+331105.1 it is 
concentrated along the bar. The SFR is highest for SDSS 153821.22+331105.1; it is probably triggered by gas flowing along the bar in the
galaxy although the emission is surprisingly faint in the disk. \\
{\bf 5.~Are these galaxies interacting?~:~}SBS~1325+597 has a disturbed optical and HI morphology. SDSS 143052.33+551440.0
also has a disturbed optical morphology and its CO profile is not symmetric about its systemic velocity.  SDSS 153821.22+331105.1 has
a bar that may have been triggered by an interaction. CG~598 appears to be 
accreting a companion in its SDSS g image and its CO profile is also asymmetric about its systemic velocity. Thus all the detected galaxies
in our sample show some signs of interaction.  

\vspace{-1mm}
\section{Implications}

The main implications of this study is that cold gas and star formation are present in voids, even though the overall environment is 
underdense. Our sample galaxies also show disturbed morphologies, possibly due to interactions with companion galaxies. Our results can be 
understood in the hierarchical picture of void evolution, in which voids merge leaving behind a filamentary substructure. Galaxies grow 
along these filaments and in clusters where filaments intersect (e.g. Sahni et al. 1994; Sheth \& van de Weygaert 2004; Cautun et al. 2014). 
The presence of both molecular gas and star formation in void galaxies indicates that they are probably evolving within this void substructure.  
Gas flowing along the filaments can be accreted by these galaxies and will contribute to the accumulation of neutral gas in their disks. High 
enough gas surface densities will result in the onset of star formation, leading to galaxy evolution within the void environment. 

\vspace{2mm}
\noindent
{\bf References}\\
Beygu, B.; Kreckel, K.; van de Weygaert, R.; et al. 2013, AJ, 145, 120\\
Cautun, M.; van de Weygaert, R.; Jones, B. J. T. et al. 2014, MNRAS, 441, 2923 \\
Cruzen, Shawn; Wehr, Tara; Weistrop, Donna et al. 2002, AJ, 123, 142\\
Karachentsev, V. E.; Karachentsev, I. D.; Richter, G. M. 1999 A\&AS, 135, 221\\
Kreckel, K.; Platen, E.; Aragón-Calvo, M. A.; et al. 2012, AJ, 144, 16 \\
Kreckel, K.; Platen, E.; Aragón-Calvo, M. A.; et al. 2011, AJ, 141, 4\\
Nakajima, Taku; Sakai, Takeshi; Asayama, Shin'ichiro; et al. 2008, PASJ, 60, 435 \\
Sage, L. J.; Weistrop, D.; Cruzen, S.; Kompe, C. 1997, AJ, 114, 1753\\	
Sahni, Varun; Sathyaprakah, B. S.; Shandarin, Sergei F. 1994, ApJ, 431, 20\\
Sheth, Ravi K.; van de Weygaert, Rien  2004, MNRAS, 350, 517\\
Solomon, P. M.; Vanden Bout, P. A. 2005, ARA\&A, 43, 677\\
Szomoru, Arpad; van Gorkom, J. H.; Gregg, Michael D. 1996, AJ, 111, 2141

\vspace{1mm}
\noindent
{\bf Acknowledgements}\\
This paper was based on observations at the Nobeyama Radio Observatory (NRO) which 
is a branch of the National Astronomical Observatory of Japan, National Institutes
of Natural Sciences. The optical observations were done at the Indian Optical Observatory
(IAO) at Hanle. We thank the staff of IAO, Hanle and CREST, Hosakote, that made these
obervations possible. This research has made use of the NASA/IPAC Extragalactic Database (NED). 

\end{document}